\documentstyle[twoside]{article}

\catcode`\@=11
\long\def\@makefntext#1{
\protect\noindent \hbox to 3.2pt {\hskip-.9pt  
$^{{\eightrm\@thefnmark}}$\hfil}#1\hfill}		

\def\thefootnote{\fnsymbol{footnote}}
\def\@makefnmark{\hbox to 0pt{$^{\@thefnmark}$\hss}}	
	
\def\ps@myheadings{\let\@mkboth\@gobbletwo
\def\@oddhead{\hbox{}
\rightmark\hfil\eightrm\thepage}   
\def\@oddfoot{}\def\@evenhead{\eightrm\thepage\hfil
\leftmark\hbox{}}\def\@evenfoot{}
\def\sectionmark##1{}\def\subsectionmark##1{}}

\oddsidemargin=\evensidemargin
\renewcommand{\thefootnote}{\fnsymbol{footnote}}

\newcounter{sectionc}\newcounter{subsectionc}\newcounter{subsubsectionc}
\renewcommand{\section}[1] {\vspace{12pt}\addtocounter{sectionc}{1} 
\setcounter{subsectionc}{0}\setcounter{subsubsectionc}{0}\noindent 
	{\tenbf\thesectionc. #1}\par\vspace{5pt}}
\renewcommand{\subsection}[1] {\vspace{12pt}\addtocounter{subsectionc}{1} 
	\setcounter{subsubsectionc}{0}\noindent 
	{\bf\thesectionc.\thesubsectionc. {\kern1pt \bfit #1}}\par\vspace{5pt}}
\renewcommand{\subsubsection}[1] {\vspace{12pt}\addtocounter{subsubsectionc}{1}
	\noindent{\tenrm\thesectionc.\thesubsectionc.\thesubsubsectionc.
	{\kern1pt \tenit #1}}\par\vspace{5pt}}
\newcommand{\nonumsection}[1] {\vspace{12pt}\noindent{\tenbf #1}
	\par\vspace{5pt}}

\newcounter{appendixc}
\newcounter{subappendixc}[appendixc]
\newcounter{subsubappendixc}[subappendixc]
\renewcommand{\thesubappendixc}{\Alph{appendixc}.\arabic{subappendixc}}
\renewcommand{\thesubsubappendixc}
	{\Alph{appendixc}.\arabic{subappendixc}.\arabic{subsubappendixc}}

\renewcommand{\appendix}[1] {\vspace{12pt}
        \refstepcounter{appendixc}
        \setcounter{figure}{0}
        \setcounter{table}{0}
        \setcounter{lemma}{0}
        \setcounter{theorem}{0}
        \setcounter{corollary}{0}
        \setcounter{definition}{0}
        \setcounter{equation}{0}
        \renewcommand{\thefigure}{\Alph{appendixc}.\arabic{figure}}
        \renewcommand{\thetable}{\Alph{appendixc}.\arabic{table}}
        \renewcommand{\theappendixc}{\Alph{appendixc}}
        \renewcommand{\thelemma}{\Alph{appendixc}.\arabic{lemma}}
        \renewcommand{\thetheorem}{\Alph{appendixc}.\arabic{theorem}}
        \renewcommand{\thedefinition}{\Alph{appendixc}.\arabic{definition}}
        \renewcommand{\thecorollary}{\Alph{appendixc}.\arabic{corollary}}
        \renewcommand{\theequation}{\Alph{appendixc}.\arabic{equation}}
        \noindent{\tenbf Appendix \theappendixc #1}\par\vspace{5pt}}
\newcommand{\subappendix}[1] {\vspace{12pt}
        \refstepcounter{subappendixc}
        \noindent{\bf Appendix \thesubappendixc. {\kern1pt \bfit #1}}
	\par\vspace{5pt}}
\newcommand{\subsubappendix}[1] {\vspace{12pt}
        \refstepcounter{subsubappendixc}
        \noindent{\rm Appendix \thesubsubappendixc. {\kern1pt \tenit #1}}
	\par\vspace{5pt}}

\newcommand{\textlineskip}{\baselineskip=13pt}
\newcommand{\smalllineskip}{\baselineskip=10pt}

\def\eightcirc{
\begin{picture}(0,0)
\put(4.4,1.8){\circle{6.5}}
\end{picture}}
\def\eightcopyright{\eightcirc\kern2.7pt\hbox{\eightrm c}} 

\newcommand{\copyrightheading}[1]
	{\vspace*{-2.5cm}\smalllineskip{\flushleft
	{\footnotesize International Journal of Modern Physics A, #1}\\
	{\footnotesize $\eightcopyright$\, World Scientific Publishing
	 Company}\\
	 }}

\def\abstracts#1#2#3{{
	\centering{\begin{minipage}{4.5in}\baselineskip=10pt\footnotesize
	\parindent=0pt #1\par 
	\parindent=15pt #2\par
	\parindent=15pt #3
	\end{minipage}}\par}} 

\newcommand{\bibit}{\nineit}

\renewenvironment{thebibliography}[1]
	{\frenchspacing
	 \ninerm\baselineskip=11pt
	 \begin{list}{\arabic{enumi}.}
	{\usecounter{enumi}\setlength{\parsep}{0pt}
	 \setlength{\leftmargin 12.7pt}{\rightmargin 0pt} 
	 \setlength{\itemsep}{0pt} \settowidth
	{\labelwidth}{#1.}\sloppy}}{\end{list}}

\newcounter{itemlistc}
\newcounter{romanlistc}
\newcounter{alphlistc}
\newcounter{arabiclistc}
\newenvironment{itemlist}
    	{\setcounter{itemlistc}{0}
	 \begin{list}{$\bullet$}
	{\usecounter{itemlistc}
	 \setlength{\parsep}{0pt}
	 \setlength{\itemsep}{0pt}}}{\end{list}}

\newcommand{\fcaption}[1]{
        \refstepcounter{figure}
        \setbox\@tempboxa = \hbox{\footnotesize Fig.~\thefigure. #1}
        \ifdim \wd\@tempboxa > 5in
           {\begin{center}
        \parbox{5in}{\footnotesize\smalllineskip Fig.~\thefigure. #1}
            \end{center}}
        \else
             {\begin{center}
             {\footnotesize Fig.~\thefigure. #1}
              \end{center}}
        \fi}

\newcommand{\tcaption}[1]{
        \refstepcounter{table}
        \setbox\@tempboxa = \hbox{\footnotesize Table~\thetable. #1}
        \ifdim \wd\@tempboxa > 5in
           {\begin{center}
        \parbox{5in}{\footnotesize\smalllineskip Table~\thetable. #1}
            \end{center}}
        \else
             {\begin{center}
             {\footnotesize Table~\thetable. #1}
              \end{center}}
        \fi}

\def\@citex[#1]#2{\if@filesw\immediate\write\@auxout
	{\string\citation{#2}}\fi
\def\@citea{}\@cite{\@for\@citeb:=#2\do
	{\@citea\def\@citea{,}\@ifundefined
	{b@\@citeb}{{\bf ?}\@warning
	{Citation `\@citeb' on page \thepage \space undefined}}
	{\csname b@\@citeb\endcsname}}}{#1}}

\newif\if@cghi
\def\cite{\@cghitrue\@ifnextchar [{\@tempswatrue
	\@citex}{\@tempswafalse\@citex[]}}
\def\citelow{\@cghifalse\@ifnextchar [{\@tempswatrue
	\@citex}{\@tempswafalse\@citex[]}}
\def\@cite#1#2{{$\null^{#1}$\if@tempswa\typeout
	{IJCGA warning: optional citation argument 
	ignored: `#2'} \fi}}

\def\pmb#1{\setbox0=\hbox{#1}
	\kern-.025em\copy0\kern-\wd0
	\kern.05em\copy0\kern-\wd0
	\kern-.025em\raise.0433em\box0}

\def\fnt#1#2{\footnotetext{\kern-.3em
	{$^{\mbox{\scriptsize #1}}$}{#2}}}

\def\fpage#1{\begingroup
\voffset=.3in
\thispagestyle{empty}\begin{table}[b]\centerline{\footnotesize #1}
	\end{table}\endgroup}

\def\runninghead#1#2{\pagestyle{myheadings}
\markboth{{\protect\footnotesize\it{\quad #1}}\hfill}
{\hfill{\protect\footnotesize\it{#2\quad}}}}
\font\tenrm=cmr10
\font\tenit=cmti10 
\font\tenbf=cmbx10
\font\bfit=cmbxti10 at 10pt
\font\ninerm=cmr9
\font\nineit=cmti9

\font\eightrm=cmr8

\def\qed{\hbox{${\vcenter{\vbox{			
   \hrule height 0.4pt\hbox{\vrule width 0.4pt height 6pt
   \kern5pt\vrule width 0.4pt}\hrule height 0.4pt}}}$}}

\renewcommand{\thefootnote}{\fnsymbol{footnote}}	

\newfont{\frak}{eufm10}
\newfont{\extra}{msbm10}

\newcommand{\fra}[1]{\mbox{\frak #1}}
\newcommand{\extr}[1]{\mbox{\extra #1}}

\def\beq{\begin{equation}}
\def\eeq{\end{equation}}
\def\bea{\begin{eqnarray}}
\def\eea{\end{eqnarray}}

\def\om{\Omega}
\def\k{\omega}
\def\s{so}

\def\R{\extr{R}}

\def\J{\Omega}

\def\kin{\k_1,\k_2,\dots,\k_N}

\def\kiin{\k_2,\dots,\k_N}

\def\Gc{\Gamma^{(m)}}
\def\CKspace#1{{\cal S}^{(#1)}}

\def\cext{\alpha}

\def\cextiif{\alpha^{F}}
\def\cextiil{\alpha^{L}}
\def\cextiii{\beta}

\def\Cos{{\mbox{$\ \!\rm C$}}}        
\def\Sen{{\mbox{$\ \!\rm S$}}}        
\def\Ver{{\mbox{$\ \!\rm V \!$}}}        

\def\kp{\k_1}                 
\def\kl{\k_2}                 
 
\def\sa{\Sen_{\kp}(a)\,}      
\def\va{\Ver_{\kp}(a)\,}      

\def\sA{\Sen_{\kl}(A)\,}      
\def\vA{\Ver_{\kl}(A)\,}      

\def\+{\!+\!}
\def\-{\!-\!}
\def\={\!\!=\!\!}

\def\comment#1{\relax}

\begin{document}

\runninghead{Geometries of orthogonal groups and their contractions $\ldots$} 
{Geometries of orthogonal groups and their contractions  $\ldots$}

\normalsize\textlineskip
\thispagestyle{empty}
\setcounter{page}{1}

\copyrightheading{}			

\vspace*{0.88truein}

\fpage{1}
\centerline{\bf GEOMETRIES OF ORTHOGONAL GROUPS}
\vspace*{0.035truein}
\centerline{\bf AND THEIR CONTRACTIONS:}
\vspace*{0.035truein}
\centerline{\bf A UNIFIED CLASSICAL DEFORMATION VIEWPOINT}
\vspace*{0.37truein}
\centerline{\footnotesize    M. SANTANDER}
\vspace*{0.015truein}
\centerline{\footnotesize\it Departamento de F\'{\i}sica Te\'orica, 
                             Universidad de Valladolid}
\baselineskip=10pt
\centerline{\footnotesize\it E-47011 Valladolid, Spain}
\vspace*{10pt}
\centerline{\footnotesize    F. J. HERRANZ}
\vspace*{0.015truein}
\centerline{\footnotesize\it Departamento de F\'{\i}sica, E.U. Polit\'ecnica 
                             Universidad de Burgos}
\baselineskip=10pt
\centerline{\footnotesize\it E-09006, Burgos, Spain}
\vspace*{0.225truein}
\bigskip

\medskip

\vspace*{0.21truein}
\abstracts{The geometries of spaces having as groups the real orthogonal
groups and some of their contractions are described from a common point of
view. Their central extensions and Casimirs are explicitly given. An
approach to the trigonometry of their spaces is also advanced.}{}{}


\vspace*{1pt}\textlineskip	

\textheight=7.8truein
\setcounter{footnote}{0}
\renewcommand{\thefootnote}{\alph{footnote}}


\section{Aim and Outline}	
\vspace*{-0.5pt}
\noindent

The general aim of this paper is to describe a particular case of a 
classical scheme which involves a whole class of spaces, and  geometries
associated to a family of Lie groups. At all different levels of this
scheme, either the spaces, the Lie groups or their Lie algebras are related
among themselves by contractions, yet their properties can be dealt with
in a completely unified way. The family we will consider here comprises
the classical real geometries of spaces with a projective metric
(Cayley-Klein or CK geometries)
\cite{SommerYRY}, together with their Lie groups and Lie algebras. These
algebras will be called here orthogonal CK algebras, as they include all
simple orthogonal real Lie algebras $so(p,q)$, as well as many Lie
algebras of great physical relevance, (Poincar\'e, Galilei, Euclidean,
etc) which are obtained by different contractions from $so(p,q)$. All
two-point homogeneous symmetric spaces of real type (the Riemannian spaces
of constant curvature) appear related to CK algebras, but pseudoRiemannian
and degenerate Riemannian spaces (which {\it strictu senso} are not
two-point homogeneous) are included as well; indeed this class of spaces
is more natural than the class of two-point homogeneous spaces as usually
considered in the literature. 

The orthogonal CK scheme is one of the several (though in reduced number)
possible schemes of a similar kind. Each include some simple real Lie
algebras, as well as some non-simple contracted algebras, which  are
however `near' to the simple ones so that most properties of simple
algebras (geometries, groups), when suitably reformulated, still survive
for them; this makes `quasi-simple' an apt name for these algebras
\cite{Ros}. The algebras we will study here can therefore be called also
`quasi-orthogonal'; a further reason to study first this family is that
all other families always include a orthogonal CK subalgebra. 

All orthogonal CK algebras allow many of its properties to be simultaneouly
studied. As two recent examples, we give here the general result for the
second cohomology group of CK algebras (Section 3) and general expressions
for all their Casimir operators (Section 4). The study of the geometry of
homogeneous spaces associated to CK algebras, sketched in Section 5, is the
paradigm of this simultaneous approach. The paper is meant as an overview
of the CK approach, including results and topics under current
development. Another part of this project, which deals with expansions
(somehow the opposite proccess to contractions) is covered in another
paper in this volume \cite{NNSexpan}.


\section{The orthogonal CK family}	
\vspace*{-0.5pt}
\noindent

Let $\k_1,\dots,\k_N$ be $N$ real numbers. For $a,b=0,1,\dots, N, \ a<b$,
let us denote
$\k_{ab}:=\k_{a\+1}\k_{a\+2}\ldots\k_b=\prod_{i=a+1}^{b}\k_{i}$ (thus,
$\k_{ab}\k_{bc}=\k_{ac}$), and consider a family (parametrized by the
constants $\k_i$) of real Lie algebras with $N(N\+1)/2$ generators
$\om_{ab}, \ a<b$, whose Lie brackets are given by 
\beq
[\om_{ab}, \om_{ac}] =  \k_{ab}\om_{bc}, \quad
[\om_{ab}, \om_{bc}] =        -\om_{ac}, \quad
[\om_{ac}, \om_{bc}] =  \k_{bc}\om_{ab}, \quad  a<b<c,
\label{ba}  
\eeq
and commutators involving four different indices are zero. 
These algebras are called orthogonal CK Lie algebras (or quasi-orthogonal
algebras) and appear as a natural  particular subfamily of all the graded
contractions from the Lie algebra $so(N\+1)$ \cite{HMOSHerSanGSGC} 
corresponding to a
${\extr Z}_2^{\otimes N}$ grading of $so(N\+1)$; this grading is related
to a set of $N$ commuting involutive automorphisms $I_{(1)}, I_{(2)},
\dots I_{(N)}$ of the algebra (\ref{ba}). A natural labelling for these
algebras is
$\s_{\kin} (N\+1)$, a notation which makes explicit the parameters $\k_i$
and generalizes the standard
$\s(N\+1)
\equiv \s_{1, 1, \dots, 1}(N\+1)$. The CK algebras  have a (vector)
representation by
$(N\+1)\times (N\+1)$ real matrices:
$ \om_{ab}=-\k_{ab}e_{ab}+e_{ba} $ where $e_{ab}$ is the matrix with a
single  non-zero entry, 1,  in the row $a$, column $b$. Since each
coefficient $\k_i$ may take positive, negative or zero values, and by
means of a simple rescaling of the initial generators can be reduced to
the standard values of $1$, $-1$ or $0$, it is clear that the family
$so_{\k_1,\dots,\k_N}(N\+1)$ includes $3^N$ Lie algebras, which are 
different as graded contractions, even if two of them can still be
isomorphic.

When all the $\k_i$ are non-zero but some of them are negative, each
algebra is isomorphic to a certain pseudo-orthogonal algebra $\s (p,q)$
$(p+q=N\+1,\, p\ge q > 0)$. When however a given  constant $\k_i=0$, the
algebra has a semidirect structure. These are obtained from $\s (p,q)$
through a sequence of In\"on\"u-Wigner (IW) contractions, and a
distinctive  trait of the CK scheme is the ability to describe the
resulting  contracted algebras by simply setting a constant equal to
zero, $\k_i=0$. When several constants
$\k_i$ are equal to zero, the resulting algebra has several  semidirect
splittings (and hence several inhomogeneous structures) into an abelian
subalgebra and an eventual direct sum of CK algebras in lower dimensions. 

For any
$m$, $(m=1, \dots, N)$, let ${\fra h}^{(m)}$ denote the subalgebra
generated by those $\J_{ab}$ where the indices $a,b$ satisfy either $b <
m$ or $a \geq m$.  A complement for ${\fra h}^{(m)}$ is the vector
subspace ${\fra p}^{(m)}$ (not always a subalgebra) spanned by the
elements $\J_{ab}$ with $a < m$ and $b \geq m$.  The structure of the
subalgebras ${\fra h}^{(m)}$ and of the vector subspace ${\fra p}^{(m)}$
of the Lie algebra $\s_{\kin} (N\+1)$ can be graphically displayed by
arranging the generators of $so(N\+1)$ in the form of a  triangle

\medskip

\noindent\hskip 1truecm
\begin{tabular}{cccc|cccc}
$\J_{01} $&$ \J_{02} $& $ \ldots $&$ \J_{0\,m\-1} $&$ \J_{0m} $
&$ \J_{0\,m\+1}$& $ \ldots $&$ \J_{0N} $  \\
 &$ \J_{12} $& $ \ldots $&$ \J_{1\,m\-1} $&$ \J_{1m} $
&$ \J_{1\,m\+1}$& $ \ldots $&$ \J_{1N} $  \\
 & & $ \ddots $&$ \vdots $&$ \vdots$
&$  \vdots$& $   $&$ \vdots $  \\
 & &  &$ \J_{m\-2\,m\-1} $&$ \J_{m\-2\,m} $
&$ \J_{m\-2\,m\+1}$& $ \ldots $&$ \J_{m\-2\,N} $  \\
 & &  & &$ \J_{m\-1\,m} $
&$ \J_{m\-1\,m\+1}$& $ \ldots $&$ \J_{m\-1\,N} $  \\
\cline{5-8}
 & &  &   \multicolumn{2}{c}{\,}
&$ \J_{m\,m\+1}$& $ \ldots $&$ \J_{m\,N} $  \\
 & &  & \multicolumn{2}{c}{\,}
& & $ \ddots $&$ \vdots $  \\
 & &  & \multicolumn{2}{c}{\,}
& &  &$ \J_{N-1\,N} $
\end{tabular}

\medskip

\noindent
We see that the subspace
${\fra p}^{(m)}$ is spanned by those $m(N\+1\-m)$ generators in the
rectangle whose bottom left corner is $\J_{m\-1\, m}$, the  left and down
triangles  corresponding respectively to the subalgebras 
$\s_{\k_1 \ldots \k_{m\-1}}(m)$ and 
$\s_{\k_{m\+1}\ldots \k_N}(N\+1\-m)$ respectively, whose direct sum is
the subalgebra  ${\fra h}^{(m)}$. In fact the action of the involutions
$I_{(m)}$ on the generators have ${\fra p}^{(m)}$ and ${\fra h}^{(m)}$ as
the antiinvariant and invariant subspaces, and  the decomposition
$\s_{\kin} (N\+1) = {\fra p}^{(m)} \oplus {\fra h}^{(m)}$ is a
Cartan-like decomposition, each of which has an associated IW
contraction, defined as the
$\epsilon\to 0$ limit of the  CK algebra transformation: 
\beq 
\begin{array}{l}
\extr X \mapsto\Gc (\extr X)= \left\{\begin{array}{lc}
\  \extr X \quad &\mbox \ \mbox{if \quad }\ \extr X\in \fra h^{(m)}\\
\epsilon   \extr X \quad &\ \mbox{if \quad}\ \extr X \in \fra p^{(m)}
\end{array}
\right. \qquad   m=1,\dots,N. 
\label{bc}
\end{array}
\eeq
It is clear that under the contraction $\Gc$, the algebra
$so_{\kin}(N\+1)$ goes to another algebra in the CK family, with the same
values of the $\k_i$ constants except for the new $\k_{m} =0$.  In the
triangular arrangement of generators, the generators in the rectangle 
are abelianized, while commutators where at least a generator is outside
this rectangle remain unchanged. 


\section{Central extensions}	
\vspace*{-0.5pt}
\noindent

Any central extension  $\overline{so}_{\k_1,\dots,\k_N}(N\+1)$ of the
algebra $so_{\k_1,\dots,\k_N}(N\+1)$ by the one-dimensional algebra of 
generator $\Xi$
will have generators $(\om_{ab},\Xi)$ and commutators: 
\beq
[\om_{ab},\om_{cd}] ={\sum_{{\scriptstyle i,j=0 \  i<j}}^N} 
            C_{ab,cd}^{ij}\om_{ij}+ \cext_{ab,cd} \Xi,\quad
[\Xi,\om_{ab}]=0, 
\label{ca}
\eeq
where $C_{ab,cd}^{ij}$ are those in eq. (1).  The extension is determined
by a two-cocycle (whose components in the $\om_{ab}$ basis are
$\cext_{ab,cd}$), and two-cocycles differing by a two-coboundary
correspond to equivalent extensions, which are therefore classified by
the second cohomology group of the algebra. The problem of determining
the central extensions of the algebra (\ref{ba}) has therefore two
stages. First, find the possible cocycles (the two-cocycle condition
reducing to the requirement that (\ref{ca}) is a Lie algebra). Second,
determine when a two-cocycle is trivial (i.e., classify these into
cohomology classes). 

The general solution of these two problems can be stated in the following
\cite{AHPS}: 

\vspace*{12pt}
\noindent
{\bf Theorem~1:} The independent non-zero commutators
of any central extension $\overline{so}_{\k_1,\dots,\k_N}(N\+1)$ of the
CK Lie algebra
${so}_{\k_1,\dots,\k_N}(N\+1)$ can be written as:
\beq
\begin{array}{ll}
\!\! [\om_{ab}, \om_{bc}] = - \om_{ac}                       &\  \cr
\!\! [\om_{ab}, \om_{a\,b+1}] = \k_{ab} \om_{b\,b+1}+\k_{a\,b-1}
                             \cextiif_{b\,b+1}\Xi          &\
\!\! [\om_{ab}, \om_{ac}] = \k_{ab} \om_{bc}
                             \quad \mbox{\rm for}\  c > b+1     \cr
\!\! [\om_{ac}, \om_{a+1\,c}] = \k_{a+1\,c} \om_{a\,a+1}+\k_{a+2\,c}
                             \cextiil_{a\,a+1}\Xi          &\
\!\! [\om_{ac}, \om_{bc}] = \k_{bc}\om_{ab}
                             \quad \mbox{\rm for}\ b > a+1      \cr
 \multicolumn{2}{l}
 { [\om_{a\,a+1}, \om_{c\,c+1}] = \cextiii_{a+1\,c+1} \Xi  \qquad
  \qquad\qquad [\om_{a\,a+2}, \om_{a+1\,a+3}] = -\k_{a+2} 
  \cextiii_{a+1\,a+3} \Xi }
\end{array}
\label{cb}
\eeq
where any $\k_{aa}$ with two equal indices should be understood as
$\k_{aa}:=1$.  The extension is completely described by a number of
extension coefficients:

\noindent
 $\bullet$   Two {\it single} coefficients, $\cextiil_{01}$, 
$\cextiif_{N-1\,N}$. The extension  $\cextiil_{01}$ (resp.
$\cextiif_{N-1\,N}$) is non-trivial if $\k_2=0$ (resp $\k_{N-1}=0$) and
is trivial otherwise.   

\noindent
 $\bullet$   $(N-2)$ {\it pairs}, $\cextiif_{12},\cextiil_{12}$,
\dots , 
$\cextiif_{N-2\, N-1},\cextiil_{N-2\,N-1}$. Each pair of coefficients
must satisfy 
$\k_{a+3}\cextiif_{a+1\,a+2} =
\k_{a+1}\cextiil_{a+1\,a+2}$. The two extensions corresponding to the
pair 
$\cextiif_{a+1\,a+2}$ and $\cextiil_{a+1\,a+2}$ are both non-trivial when
$\k_{a+1} = 0$ and  $\k_{a+3} = 0$. The two two-cocycles are
simultaneously trivial otherwise.

\noindent
$\bullet$  $(N-2)$ extension coefficients $\cextiii_{13},
\cextiii_{24},
\dots, \cextiii_{N-2\,N}$.  The coefficient $\cextiii_{b+1\,b+3}$ 
satisfies $ \omega \cextiii_{b+1\,b+3} = 0
\quad\hbox{for}\quad
\omega=\k_{b},\; \k_{b+1}\k_{b+2},\; \k_{b+2}\k_{b+3}, \; \k_{b+4}$ where
when either
$b=0$ or $b=N-3$ the first or last conditions, which would read 
$\k_{0}\cextiii=0$ or $\k_{N\+1}\cextiii=0$ are not present. The extension
corresponding to any of these non-zero coefficients is always non-trivial.

\noindent
 $\bullet$  $(N-2)(N-3)/2$ extension coefficients $\cextiii_{14},
\cextiii_{15}, \dots, \cextiii_{1N}, \cextiii_{25}, \dots,
\cextiii_{2N},
\dots, \cextiii_{N-3\,N}$ whose indices differ by more than two. The
coefficient $\cextiii_{b+1\,d+1}$ satisfies 
$  \omega\cextiii_{b+1\,d+1}  =0\quad \hbox{for}\quad
\omega=\k_{b},\; \k_{b+2},\; \k_{d},\; \k_{d+2} $ with similar
restrictions as to the actual presence of the equations involving the
non-existent constants $\k_{0}$ or
$\k_{N\+1}$. The extension corresponding to any of these non-zero
coefficients is always non-trivial. 


\section{Casimir Operators}	
\vspace*{-0.5pt}
\noindent

When all $\k_i$ are different from zero, 
${so}_{\k_1,\dots,\k_N}(N\+1)$ is a  simple algebra, whose rank is
$l=[\frac{N\+1}{2}]$. The dimension of the center of its universal
enveloping algebra equals the rank $l$ of the algebra, and is generated
by a set of  homogeneous polynomials (Casimir operators) of orders $2, 4,
\dots, [\frac{N}{2}]$, and an additional Casimir of order $l$ when
$N\+1$ is even. Upon any contraction $\k_m \to 0$, and after a suitable
rescaling, these Casimirs go into non-trivial Casimir operators for the
contracted algebra. Furthermore, these are a complete set of Casimirs for
CK algebras. This provides another justification for the name
`quasi-simple'  given to its members. 

We now give \cite{HerSanCasimirs} the expressions for the
$[\frac{N\+1}{2}]$ Casimir operators in the general CK Lie algebra
${so}_{\k_1,\dots,\k_N}(N\+1)$. We do this by using an approach where the
behaviour of Casimirs under the family of contractions $\k_m \to 0$ is
transparent. 

Starting from the generators of the Lie algebra
${so}_{\k_1,\dots,\k_N}(N\+1)$, we define some elements in its universal
enveloping algebra, $W_{ab}, W_{a_1a_2b_1b_2}, \dots$. The recursive
definition starts from $W_{ab} := \om_{ab}, (a<b)$ and further $W$ symbols
with four indices $W_{a_1a_2b_1b_2}$ (where $a_1 < a_2 < b_1 < b_2$ ), six
indices  \dots, until symbols with $2l$ indices $W_{a_1a_2\dots
a_lb_1b_2\dots b_l}$  (with $a_1 < a_2 < \dots < a_l < b_1 < b_2 < \dots
<b_l$), are given in terms of those with two less indices. For $s=2,
\dots, l$ the
$2s$-index symbol $W$ is given in terms of the $(2s\-2)$-index symbol $W$
by means of:
\bea
&& W_{a_1a_2\dots a_sb_1b_2\dots b_s} =
  {\sum_{\mu=1}^{s}} (-1)^{\mu} \om_{a_\mu b_s}
  W_{\{a_1a_2\dots a_sb_1b_2\dots b_s\} - \{a_\mu b_s\}} \cr
&& \qquad\qquad\qquad\qquad
  + \ {\sum_{\nu=1}^{s-1}} (-1)^{s+\nu} \k_{a_sb_\nu} \om_{b_\nu b_s}
        W_{\{a_1a_2\dots a_sb_1b_2\dots b_s\} - \{b_\nu b_s\}}.
 \label{da}
\eea
where the $W$ symbols in the right hand side of the equation have $2s\-2$
indices, those obtained by {\it removing} the two indices $\{a_\mu b_s\}$
from the set of
$2s$ indices ${\{a_1a_2\dots a_sb_1b_2\dots b_s\}}$. In terms of these
symbols, the Casimirs are given by:

\vspace*{12pt}
\noindent
{\bf Theorem~2:} The $l=[\frac{N\+1}{2}]$ independent polynomial Casimir
operators of the CK Lie algebra ${so}_{\k_1,\dots,\k_N}(N\+1)$ can be
written as:
\begin{itemlist}
\item $[\frac{N}{2}]$ invariants ${\cal C}_s$, $s=1, \dots, [\frac{N}{2}]$
      of order $2s$:
\beq
{\cal C}_s = \!\!\!\!\!\!\!\!\!\!\!\!\!\!\!\!\!\!\!\!\!
             {\sum_{a_1a_2\dots a_sb_1b_2\dots b_s =0 \atop 
                  a_1<a_2<\dots <a_s<b_1<b_2<\dots<b_s }^{N}}
             \!\!\!\!\!\!\!\!\!\!\!\!\!\!\!\!\!\!\!\!\!
       \k_{0a_1}\k_{1a_2}  \dots  \k_{s\-1 a_s}
       \k_{b_1 N\-s\+1}\k_{b_2 N\-s\+2}  \dots  \k_{b_s N} 
       W_{a_1a_2\dots a_sb_1b_2\dots b_s }^2 
\label{db}
\eeq
\item When $N\+1$ is even, there is an extra Casimir ${\cal C}$ of
      order $l=\frac{N\+1}{2}$, $ {\cal C} = W_{012 \dots N} $.
\end{itemlist}

Again in these expresions any $\k_{aa}$ should be understood as
$\k_{aa}:=1$. It is easy to see that even in the most contracted CK
algebra, the flag space algebra, ${so}_{0,\dots,0}(N\+1)$, these Casimirs
are not trivial. In fact, the term in eq. (\ref{db}) with the $W$ symbol
whose first group of $s$ indices are consecutive and start from $0$, and
whose last group of $s$ indices are also consecutive and end with $N$,
$W_{012\dots s\-1 N\-s\+1
\dots N\-2 N\-1 N}^2$ is the only term whose $\k$ factor is equal to 1, 
and therefore the only which survives in the Casimir ${\cal C}_s$ for the
${so}_{0,\dots,0}(N\+1)$ algebra. 


\section{The CK scheme of homogeneous spaces}	
\vspace*{-0.5pt}
\noindent
 
An aggregate of $N$ homogeneous symmetric spaces can be allocated to each
algebra $\s_{\kin} (N\+1)$. If the group obtained by exponentiation of
the generators $\J_{ab}$ is naturally denoted as
$SO_{\kin}(N\+1)$, then the subgroup generated by ${\fra h}^{(m)}$ is
easily shown to be $SO_{\k_1, \dots, \k_{m\-1}}(m) \otimes SO_{\k_{m\+1},
\dots, \k_N}(N\+1\-m)$. Therefore, for each $m=1, \dots, N$ we can
construct the coset space $\CKspace{m} := SO_{\kin} (N\+1) / [SO_{\k_1,
\dots,
\k_{m\-1}}(m) \otimes SO_{\k_{m\+1}, \dots,\k_N}(N\+1\-m)]$, of dimension
$m(N\+1\-m)$, which is a homogeneous space for the CK group $SO_{\kin}
(N\+1)$. Furthermore, the $N$ commuting involutive automorphisms
$I_{(1)} \dots I_{(N)}$ endow these coset spaces with a structure of
symmetric space, the involution $I_{(m)}$ playing the role of geodesic
reflection around the origin in $\CKspace{m}$. Of course, all these
homogeneous spaces are built from the same Lie algebra/group, with a
unified procedure, and they are interlinked, as each space can be
interpreted in terms of any other. This is in tune with a `Hilbert-like'
approach, where points lines, planes, \dots , appear with the same
footing and no preponderance is given to any of them above others. 

In terms of the triangular arrangement of generators, the $N$ possible
homogeneous spaces associated to the CK $\s_{\kin} (N\+1)$ algebra
correspond to the $N$ different rectangles that can be selected inside the
large triangle. The generators in this rectangle generate translations
along lines through the origin of the space, and the subspace ${\fra
p}^{(m)}$ is identified to the tangent space to $\CKspace{m}$ at the
origin.   

Given two different points in $\CKspace{m}$, it can be shown that the
number of functionally independent invariants under the group action is
the minimum of the two values $m, N\+1\-m$. This number will be called
here the {\it rank} of the CK homogeneous space. When all constants
$\k_i$ are positive (and therefore can be asumed to be equal to 1), the
homogeneous space is exactly the Grassmannian 
$G(m, N\+1\-m) \equiv SO(N\+1)/(SO(m) \otimes SO(N\+1\-m)$ of
$m$-planes in the $\R^{N\+1}$ ambient space; for this case (in general,
whenever all $\k_i \neq 0$), the rank of the space also coincides with
the maximal dimension of a totally geodesic and flat submanifold.


\subsection{CK geometries}
\noindent

Each CK symmetric homogeneous space $\CKspace{m}$ inherits from its
algebra/group: 
\begin{itemlist}
 \item an invariant canonical connection, associated to the symmetric
       homogeneous space structure
 \item A (possibly degenerate) main metric, coming from a suitable
       rescaling of the Killing-Cartan form.
 \item When one of the constants 
       $\k_{1},\k_{2},\dots, \k_{m\-1},\k_{m\+1},
       \dots, \k_{N}$ is equal to zero, then the space $\CKspace{m}$
       has a fibered structure. This can be considered as an invariant
       foliation, each of whose leaves carries a subsidiary metric, 
       coming again from the Killing-Cartan form through
       restriction to the leaves and suitable rescaling. 
 \item Finally, the canonical connection and the complete hierarchy of
       subsidiary metrics are compatible. 
\end{itemlist}

All the constants $\kin$ receive a geometrical meaning in the spaces
$\CKspace{m}$. The metrics in $\CKspace{m}$ depends on the constants
$\k_{1}, \dots, \k_{m\-1}, \k_{m\+1}, \dots, \k_{N}$. At the tangent
space at the origin, the `main' metric matrix has the diagonal block
structure diag$(g^{(m)}, \k_{m\-1}g^{(m)}, \k_{m\-2}\k_{m\-1}g^{(m)},
\dots,
\k_{1} \dots \k_{m\-2}\k_{m\-1}g^{(m)} )$, where each block $g^{(m)}$
itself is a diagonal matrix diag$(1, \k_{m\+1}, \k_{m\+1}\k_{m\+2},
\dots ,\k_{m\+1}\k_{m\+2}\dots \k_{N})$. When all these constants are
different from zero, the main metric in
$\CKspace{m}$ is non-degenerate; the invariant canonical connection turns
out to be the corresponding Levi-Civita metric connection, and the value
$\k_{m}$ appears as the sectional curvature of certain plane directions;
strictly speaking these spaces are not of constant curvature as this is
usually understood for (rank one) Riemannian spaces, yet this structure
is as close to constant curvature as a higher rank space can be. 

Let us note that within the CK space $\CKspace{1} \!\equiv\! SO_{\kin}
(N\+1) / SO_{\kiin}(N)$, the subalgebras ${\fra h}^{(m)}$
are identified with the isotropy subalgebras 
of a point (for $m=1$), 
of a line (for $m=2$), 
\dots , 
of a hyperplane (for $m=N$).  As ${\fra h}^{(m)}$ is the isotropy
subalgebra of the elements of the space
$\CKspace{m}$, we get an easy way of visualizing all the CK spaces:
$\CKspace{2}$ is the space of all lines in 
$\CKspace{1}$, $\CKspace{3}$ is the space of all 2-planes in 
$\CKspace{1}$, and so on. In terms of the rank 1 space 
$\CKspace{1}$ the geometrical meaning of the set of $N$ contractions
$\Gc$ is to describe the behaviour of the space around a point, a line,
\dots, a ($N\-1$)-plane (hyperplane). Remember that $\k_m$ is the
curvature of the space $\CKspace{m}$; when translated to the space
$\CKspace{1}$ the $\k_i$ describe the behaviour of the space around a
point ($\k_1$), a line ($\k_2$), \dots, an hyperplane ($\k_N$). 

The systematic use of these ideas  affords a very clear and complete
image  to visualize properties of higher rank CK spaces. Indeed the same
game can be played not only in $\CKspace{1}$ but in any $\CKspace{k}$:
elements of each space $\CKspace{m}$ appear naturally as certain
submanifolds in  $\CKspace{k}$. For example, an element (point) in
$\CKspace{1}$ appears in the space
$\CKspace{2}$ (identified to the manifold of all lines in
$\CKspace{1}$) as the submanifold of all lines which are incident to the
given point. This way, when the elements determined by each of the
isotropy subalgebras ${\fra h}^{(m)}$ ($m=1, \dots, N$) are realized in
the coset space
$\CKspace{k}$, they appear as a submanifold whose dimension is given by 
$(m-k)k$ when $k\leq m$ and by $(k-m)(N\+1\-k)$ when $m \leq k$.  


\subsection{Trigonometry in CK spaces}
\noindent

The known analogy between hyperbolical and spherical trigonometry is
related to the known duality for their homogeneous spaces. The consistent
use of the CK scheme allows to push this analogy further on. 

Three non collinear points on three lines in a CK space $\CKspace{1}$ of
any dimension always lie on a totally geodesic two dimensional
submanifold, and therefore the study of their trigonometry reduces to the
one of the two-dimensional space $\CKspace{1}$ associated to the three
dimensional CK algebra $so_{\k_1,
\k_2}(3)$. These nine essentially different spaces are labelled by the
two constants
$\k_1, \k_2$. The three most known spaces are the two-dimensional
Riemannian spaces of constant curvature $\k_1$, (the sphere, the
Euclidean plane or the hyperbolical Lobachewski plane, with constants
$\k_1, \k_2=1$). The trigonometry  of spaces with $\k_2 \leq 0$ is less
known. For $\k_1=0, \k_2=-1/c^2$, we have the Minkowski (1+1) space,
whose trigonometry was discussed by Birman and Nomizu
\cite{BirNom}. The basic trigonometric formulas for these nine spaces are
given, without proof, in the book by Yaglom\cite{Yaglom}. We show here
how to formulate trigonometry for all nine CK 2-d spaces in one stroke.
Let us consider a pure triangle in $\CKspace{1}$: three points which
determine three lines (as a model, think of a pure triangle as a triangle
with time-like sides in Minkowski space). Then the sides $a, b, c$ and
the angles $A, B, C$ (labelled as usual, and with $a$ the largest side)
can be defined as the values of the canonical parameters in the
one-parameter family of translations along one side (rotations around a
vertex), which carry one vertex to the other vertex on the side (one side
to the other side through the vertex). This definition coincides with the
one got by using the geodesic distances asociated to the canonical
metrics in the spaces of points and of lines. A striking peculiarity
which follows from this approach should be remarked: the larger angle $A$
corresponds to the external angle in Euclid. Let us call $P_a, P_b, P_c$
the generators of translations along the three sides (all three
generators conjugated to the basic translation generator $P_1 :=
\om_{01}$ in the CK algebra), and $J_A, J_B, J_C$ the generators of
rotations around the three vertices (all three conjugated to the basic
rotation generator
$J := \om_{12}$). 

All equations of trigonometry can be derived from the two conditions:
\begin{itemlist}
 \item $\exp(bP_b) \exp(-aP_a) \exp(cP_c) $ must be a rotation around
       vertex $A$ 
 \item $\exp(BJ_B) \exp(-AJ_A) \exp(CJ_C) $ must be a traslation 
       along side $a$ 
\end{itemlist}

These requirement leads to a complete set of equations relating sides and
angles, which can be adequately written in terms of the generalised
cosine $\Cos_{\k}(x)$, sine $\Sen_{\k}(x)$, and versed sine
$\Ver_{\k}(x)$ functions, which depend on a label and reduce to either
circular or hyperbolic functions when the label is $1$ or $-1$ and to the
functions $1$, $x$ and
$x^2/2$ when the label $\k=0$ \cite{QuantCK2d}. All spherical, euclidean
and hyperbolic  trigonometry can be shown to be the particular
specialization of these general equations for the cases where the
constant $\k_2=1$, while $\k_1$ is either positive, zero or negative
\cite{Tesis}. The basic relations are:
\beq
\begin{array}{ll}
& {\sa}/{\sA} = 
  {\Sen_{\kp}(b)}/{\Sen_{\kl}(B)} = 
  {\Sen_{\kp}(c)}/{\Sen_{\kl}(C)},                                   \cr 
& \Ver_{\kp}(a)\, = \Ver_{\kp}(b+c)\, - 
                  \k_2 \Ver_{\kl}(a)\, \Sen_{\kp}(b) \Sen_{\kp}(c),    \cr 
& \Ver_{\kl}(A)\, = \Ver_{\kl}(B+C)\, - 
                  \k_1 \Ver_{\kp}(a)\, \Sen_{\kl}(B) \Sen_{\kl}(C),    \cr
\end{array}
\eeq
where one can recognize the general form of the sine, cosine and dual
cosine theorems, the two last in their versed sine form. Another
remarkable consequence we get using this approach are the identities 
\beq
\begin{array}{ll}
& \exp(bP_b) \exp(-aP_a) \exp(cP_c) = \exp((B\-A\+C)J_A), \cr
& \exp(BJ_B) \exp(-AJ_A) \exp(CJ_C) = \exp((b\-a\+c)P_a), 
\end{array}
\eeq
which give the holonomies associated to the triangle, both in the space of
points and in the space of lines. 

The ordinary duality $\k_1 \to -\k_1$ is automatically contained in the
fact that all sides appear in functions with the label $\k_1$; similarly
there is another duality for angles and $\k_2$. However the most
interesting property  in this approach is the complete invariance under
the simultaneous interchange sides
$\!\leftrightarrow\!$ angles and $\k_1 \!\leftrightarrow\! \k_2$. 

When both constants $\k_1, \k_2 \to 0$, these equations reduce to:
\beq
\frac{a}{A} = \frac{b}{B} = \frac{c}{C}, 
\qquad A^2 = (B+C)^2, \qquad a^2 = (b+c)^2, \label{dd}
\eeq
which for the kinematics of the (1+1) Galilean space time are elementary 
physical properties (in particular, the two last equations are but the
velocity addition law and the absolute character of time). 

A possibility opened by this approach is to consider the trigonometrical
equations for all CK spaces, and to look how these equations depend on
the two curvatures $\k_1, \k_2$. For the `fiducial' case $\k_1\=\k_2\=0$
the basic equations are simple, linear ones. One may look to the general
equations as some {\it classical deformations} of these purely linear
ones (eq. \ref{dd}). For any equation the corresponding `deformation' 
involves two stages:  Firstly, sides and angles are replaced everywhere
they appear by a trigonometrical function whose label is the appropiate
curvature: 
$ a   \!\!\to\!\! \sa, \  
a^2 \!\!\to\!\! \va,  \ 
A   \!\!\to\!\! \sA,  \ 
A^2 \!\!\to\!\! \vA$, etc. and secondly, eventual additional terms
involving explicitly the curvatures $\k_1, \k_2$ may appear. It is
however posible to write all equations in some equivalent `minimal' way,
where curvatures $\k_1, \k_2$ do not appear explicitly, sides and angles
appear always through their sines, $\sa, \sA$, and further to that, the
deformation is described by some `missing' cosines. We say here `missing'
as when $\k=0$, all cosines equal identically to 1, so that they are
simply invisible. These ideas are worth of further development.

The approach to trigonometry we have sketched here  has been already
developped for hermitian CK spaces, and should also be extended to the
scarcely known trigonometry of higher rank spaces.
 

\nonumsection{Acknowledgements}
\noindent
M.S. would like to acknowledge the organisers of the II IWCQIS for their
invitation and hospitality during the meeting. This work has been
partially supported by a DGICYT project (PB94--1115) from Ministerio de
Educaci\'on y Ciencia de Espa\~na.


\nonumsection{References}

\end{document}